**TITLE**

# A Proof-of-Concept Study of Multitask Learning for Cranial Synthetic CT Generation Across Heterogeneous MRI Field Strengths

**RUNNING TITLE**

Cranial Synthetic CT Across MRI Field Strengths

**AUTHORS**

Zhuoyao Xin[1,2,3], Yiren Zhang[4], Christopher Wu[5], Dong Liu[6], Chunming Gu[1,2,3,7], Elena Greco[8], Erik H. Middlebrooks[8], Jun Hua[1,2*], Jia Guo[6*]

**AUTHOR AFFILIATIONS**

[1] F.M. Kirby Research Center for Brain Imaging, Kennedy Krieger Institute, Baltimore, Maryland, 21205, United States.

[2] Neurosection, Division of MR Research, Russell H. Morgan Department of Radiology and Radiological Science, Johns Hopkins University School of Medicine, Baltimore, Maryland, 21287, United States.

[3] Department of Biomedical Engineering, Johns Hopkins University, Baltimore, Maryland, 21218, United States.

[4] Department of Biomedical Engineering, Case Western Reserve University, Cleveland, Ohio, 44106-4207, United States.

[5] Department of Biomedical Engineering, Columbia University, New York, New York, 10027, United States.

[6] Department of Neuroscience, Columbia University, New York, New York, 10027, United States.

[7] Department of Radiology, Mayo Clinic, Rochester, Minnesota, 55905, United States.

[8] Neuroradiology and Neurosurgery, Mayo Clinic College of Medicine and Science, Jacksonville, Florida, 32224, United States.

**CORRESPONDING AUTHORS**

Dr. Jun Hua

Department of Radiology, Johns Hopkins University School of Medicine,
600 N. Wolfe Street, Baltimore, MD 21287, USA
Email: huajun.mri@gmail.com

Dr. Jia Guo
Department of Biomedical Engineering, Columbia University,
116th Street and Broadway, New York, NY 10027, USA
Email: jg3400@columbia.edu

**ABSTRACT**

Background

Accurate synthesis of computed tomography (CT) images from magnetic resonance imaging (MRI) is clinically valuable for cranial applications such as attenuation correction, radiotherapy planning, and image-guided interventions. However, the heterogeneity of MRI data across magnetic field strengths and pulse sequences limits the generalizability of existing methods, posing a barrier to clinical translation.

Purpose

This study aims to reformulate cranial CT synthesis as a modular, structurally coupled problem using a deep learning approach, enhancing adaptability across heterogeneous MRI conditions, including different field strengths and sequence protocols.

Methods

We implemented a cascaded multitask pipeline that jointly models skull segmentation and Hounsfield Unit (HU) regression in anatomically targeted regions. A 3D patch-based training paradigm shifts the modeling focus from global image translation to localized bone feature extraction. The backbone leverages a residual Mamba-based state space model within a 3D U-Net structure to improve spatial representation, while a Transformer U-Net serves as a widely recognized reference baseline comparison. The model was trained and evaluated in terms of multi-modal (T1-weighted and T2-FLAIR) and cross-domain transferability from a public 1.5 Tesla brain dataset (n = 37) to an independent 7 Tesla clinical brain dataset (n = 44). Performance was assessed using Dice and Jaccard indices for segmentation accuracy and mean absolute error (MAE) for HU regression.

Results

Quantitative analysis showed that our multitask pipeline, incorporating both morphological and HU map prediction stages, significantly outperformed conventional direct MRI-to-CT mapping across all metrics on the public 1.5T dataset ($p < 0.05$). Consistent performance on an external 7T clinical dataset ($p < 0.001$) further demonstrated the method's robustness and adaptability across field strengths.

Conclusion

These findings support that task-structured training for modality transformation markedly improves both accuracy and generalizability of cranial CT synthesis across heterogeneous MRI conditions. The observed consistency across field strengths validates the robustness of the proposed methodology.

## 1. INTRODUCTION

Synthetic CT (sCT) has emerged as a promising technique for generating CT-equivalent images from MRI, enabling integration of complementary modality information[1,2] for clinical applications, such as aberration correction in focused ultrasound (FUS) therapy[3,4], attenuation correction for PET/MRI–CT fusion[5–7], and image guidance for diagnostic evaluation or radiotherapy planning[8–11]. By obviating a separate CT acquisition, sCT has the potential to reduce overall acquisition time, cost, and ionizing radiation exposure, while circumventing inter-modality registration errors[12–14]. The principal technical challenge for sCT lies in the accurate depiction of cranial cortical bone[15]: conventional MRI provides little or no reliable signal in dense bone, whereas CT uniquely encodes bone mineral density and structure[16], which, for example, determine ultrasound transmission and heating in FUS[17] and help reveal cortical bone loss (osteolytic) or new bone formation (osteoblastic) in metastatic disease[18,19].

Recent advances in CT synthesis methods have moved beyond bulk density and atlas-based methods[20,21] toward predominantly deep learning-driven artificial intelligence approaches[22]. As models in computer vision continue to advance, new breakthroughs in sCT leverage discriminative architectures such as U-Net variants[10,23] and Transformer-based models[24], as well as generative frameworks including Generative Adversarial Network (GANs)[7,25] and diffusion models[26], to improve image realism and prediction accuracy. These efforts have primarily focused on improving image prediction through advancements in model capabilities. Nevertheless, sCT conversion inherently relies on the accurate extraction of bone-relevant features from MRI and the precise estimation of HU values in anatomically specific regions. Models optimized using global image-level loss functions tend to overemphasize soft tissue while underrepresent bone structures, especially in the cranial region. This imbalance can introduce training bias or require additional morphological constraints to avoid output homogeneity[27], and the need for extensive model parameter tuning and computational resources can be burdensome. Under relatively standardized imaging conditions, such limitations may be partially masked, and commercial MR-based synthetic CT solutions have demonstrated clinical feasibility in conventional workflows. However, these implementations are typically optimized for predefined scanner configurations, acquisition protocols, and conventional field strengths. In particular, variations in MRI field strength[28], from conventional imaging to ultra-high-field, introduce substantial domain shifts in contrast, artifacts, and tissue properties, potentially amplifying these unresolved challenges. As a result, systematic evaluation of sCT robustness under such heterogeneous conditions remains insufficiently characterized.

To address these challenges, we reformulate conventional modality transformation as a structurally guided, cascaded multitask framework comprising skull segmentation and anatomy-aware HU regression. This transformation introduces explicit morphological constraints and promotes task-level coupling to direct model attention toward clinically meaningful regions, thereby enhancing spatial specificity and interpretability. Designed as a versatile processing paradigm, the framework accommodates flexible backbone and output head configurations, enabling cranial sCT generation under heterogeneous MRI field strengths, with potential extension to different imaging settings. For the backbone, we adopt a 3D residual Mamba-based U-Net[29], which captures long-range dependencies more efficiently than self-attention through sequential state updates, while maintaining scalability for volumetric medical data[29–31]. For benchmarking, a Transformer U-Net[32] was also implemented as a widely accepted reference architecture.

Unlike prior studies that rely on 2D or pseudo-3D training to circumvent memory limitations[23,33], our approach preserves volumetric continuity via patch-wise 3D learning and sequential image restoration. This strategy supports flexible input sizes and mitigates slice-wise inconsistency. Additionally, most existing CT synthesis approaches focus on translating single-sequence MRI acquired at a fixed field strength—typically standard protocol sequences such as T1-weighted, T2-weighted, or Fluid Attenuated Inversion Recovery (FLAIR)[34–36], and more recently, bone-enhanced sequences such as ultra-short echo time (UTE) or zero echo time (ZTE) imaging[37–39]. To better leverage the flexibility of MRI protocol integration and evaluate model generalizability across varying acquisition conditions, we incorporate dual-modality inputs from T1-weighted and T2 FLAIR images of the human brain[40], drawn from both a public 1.5T brain dataset and independently acquired 7T clinical brain scans. While 7T MRI offers substantially higher SNR and spatial resolution, it also introduces greater B0/B1 inhomogeneity and shortened T2*[41], making it a particularly challenging but clinically relevant scenario[42]. Validation across such multimodal, multi-field-strength, and heterogeneous sources allows for more rigorous assessment of both anatomical and numerical fidelity, ultimately improving the practical utility and deployment readiness of sCT models.

## 2. METHODS

### 2.1. Data preparation

The publicly available CERMEP-IDB-MRXFDG dataset[43] was used as the primary training cohort, which included 37 healthy adult subjects (mean age 38.1 ± 11.4 years; range 23–65; 54% women). It contains matched brain images, specifically T1-weighted Magnetization-Prepared Rapid Gradient-Echo (MPRAGE), T2-weighted FLAIR MRI on a Siemens Sonata 1.5T scanner, together with a low-dose CT scan on the CT component of a Siemens Biograph mCT64 PET/CT system (only CT images were used). The T1 MPRAGE sequence was acquired in sagittal orientation with TR = 2400 ms, TE = 3.55 ms, inversion time = 1000 ms, flip angle = 8°, matrix size = 160 × 192 × 192, and voxel size = 1.2 × 1.2 × 1.2 $mm^3$. FLAIR images were acquired with TR = 6000 ms, TE = 354 ms, inversion time = 2200 ms, flip angle = 180°, matrix size = 176 × 196 × 256, and the same voxel resolution of 1.2 mm isotropic. The CT images were reconstructed with a 512 × 512 × 233 matrix and a voxel size of 0.6 × 0.6 × 1.5 $mm^3$. Background objects in the CT scans were removed using binary masks generated with FMRIB Software Library (FSL) and NiftySeg. All subject data were uniformly registered and spatially normalized to MNI space using Statistical Parametric Mapping (SPM) 12; additional post-processing details are available in the original publication. The dataset was randomly split into training, validation, and test sets (8:1:1). Both MR and CT images were subjected to a min–max normalization process to scale the values within the range of 0–1 during training. The ground truth skull labels were generated from the CT images using a fixed-intensity thresholding method (HU > 250), isolating cranial bone from adjacent gray matter, white matter, and background.

A cohort from Mayo Clinic Florida was used for external validation. Specifically, consecutive patients with both ultra-high-field 7T brain MRI and non-contrast head CT available were identified from a clinical imaging database. Consecutive enrollment was adopted to obtain a heterogeneous and representative sample of routine clinical neuroimaging cases. The dataset consisted of 44 paired brain MRI and CT scans obtained on a Siemens Magnetom Terra 7T system. Each subject underwent a sagittal Magnetization Prepared 2 Rapid Acquisition Gradient Echoes (MP2RAGE) acquisition (voxel size: 0.4 × 0.77 × 0.77 $mm^3$; TE = 2.2 ms; matrix: 384 × 296 × 296) and a T2-weighted FLAIR scan (voxel size: 0.9 × 0.44 × 0.44 $mm^3$; TE = 260 ms; matrix: 176 × 552 × 384). MP2RAGE uniform images were reconstructed following published procedures[44]. Corresponding non-contrast head CT scans were acquired on multiple routine clinical CT systems at the institute under a standardized institutional non-contrast head protocol and were reconstructed at 512 × 512 × 197 (voxel size = 0.49 × 0.49 × 1.0 $mm^3$). Rigid MRI–CT registration was performed using FSL for model fine-tuning and evaluation purposes. No additional post-processing was applied beyond the CT-based cranial mask generation.

## 2.2. Processing Framework and Network Architecture

The overall architecture of our method is illustrated in Fig. 1. To address the challenge of synthesizing CT images with anatomically accurate bone representation from MRI, we design a cascaded multitask pipeline in which skull segmentation serves as a precursor to localized HU regression. Given paired MRI and CT scans, the framework begins by extracting 3D patches based on spatial criteria. Supervised by ground-truth CT masks, the segmentation output reflects model-derived anatomical priors, which are passed into the regression module to guide voxel-wise HU prediction. Specifically, the predicted masks are subjected to geometric dilation and used to generate dynamic, structure-aware attention maps that constrain loss computation within clinically relevant regions. Although the two tasks are independently supervised, this implicit parameter sharing through structural linkage facilitates end-to-end joint training, enabling the model to synergistically learn morphological localization and intensity estimation in a unified system—directing network focus toward anatomically meaningful targets and enhancing both spatial specificity and interpretability. In parallel, the shared backbone also supports an auxiliary soft-tissue regression branch with attention emphasis on non-bone regions, and the outputs from both branches are finally fused to generate the complete sCT volume.

Specifically, for each input patch, the overall loss function is composed of a global segmentation term and a localized regression term, as formulated in Eq. (1). For morphological segmentation, a weighted sum of Dice and binary cross-entropy (BCE) losses is applied across the entire patch region to balance global structural accuracy and voxel-level boundary delineation, thereby enhancing the model's capacity for skull detection at a whole-volume scale. In contrast, the HU regression branch for bone density or soft tissue focuses on anatomically relevant regions. A dynamic attention mask is generated by applying two iterations of morphological dilation[45] applied to the predicted segmentation output from the upstream task. This mask is then used to constrain the mean squared error (MSE) computation to localized bone structures within the patch. By directing supervision toward clinically meaningful areas, this region-weighted learning strategy enhances the model's sensitivity to bone density variation. In practice, the dilated mask typically encompasses only about one-tenth of the total patch voxels, substantially reducing the computational burden during training.

$$Loss = (1-\lambda)Dice_{loss_{Global}} + \lambda BCE_{loss_{Global}} + \frac{1}{|R|}\sum_{(i,j,\kappa \in R)} MSE_{loss_{Local[i,j,\kappa]}} \tag{1}$$

Where λ balances the relative contribution of binary cross-entropy and Dice loss for global structural supervision, R denotes the set of voxels within the dilated predicted mask region, i, j and k represent the voxel's position in the patch. Patch extraction is controlled by a sampling switch that restricts the patch center to a predefined subregion. Specifically, the center is constrained to lie within a margin equal to half the patch size on each side of the image volume dimensions (H, W, D) ensuring full spatial coverage while avoiding boundary artifacts. Based on this criterion, the total number of patches used for global sCT reconstruction, as well as the number of distinct training patches that can be extracted from a single subject-can be computed as follows:

$$N_{Patch} = \left(1 + \frac{H - P_H}{S_h}\right) \times \left(1 + \frac{W - P_W}{S_w}\right) \times \left(1 + \frac{D - P_D}{S_d}\right) \tag{2}$$

Where P represents the patch size, and S is the step size along each spatial axis, which determines the degree of overlap between adjacent patches. When S exceeds the patch length in a given direction, no overlap occurs. The embedding of 3D patch extraction substantially expands the effective training dataset, enabling cubic-level sample augmentation without the need for additional data augmentation strategies. By decomposing the original global prediction task into localized structural learning, 3D training becomes feasible even with a limited number of subjects. After inference, predicted CT patches are sequentially reassembled into full volumes, with voxel-wise averaging applied in overlapping regions to improve consistency. In the reconstruction part, our image size is standardized to 207×243×226, with a cubic patch of $128^3$ voxels, a step size of 6 in each direction, yielding a total of 5670 patch samples for one subject.

Our architecture integrates multimodal 3D features through a shared encoder-decoder structure based on the U-Net framework[46], which remains a widely adopted backbone in medical imaging due to its symmetric topology, strong localization accuracy, and capability to preserve spatial resolution. To enhance long-range dependency modeling and memory efficiency in volumetric learning, we embed a residual Mamba-based state space module at the bottleneck layer of the U-Net. As a structured selective mechanism, Mamba replaces self-attention with implicit recurrence and dynamic input-conditioned transitions, enabling linear-time modeling while capturing extended contextual dependencies. This property aligns well with the nature of anatomical structure learning in high-resolution 3D data, where both spatial continuity and hierarchical context matter.  Its implicit memory control further improves training stability and robustness under limited data, addressing common limitations of attention-based models in clinical datasets. The underlying selective state-space formulation can be expressed as:

$$\mathrm{x_{t+1}} = \overline{\mathrm{A}}(\mathrm{u_t})\mathrm{x_t} + \mathrm{B}(\mathrm{u_t}) \tag{3}$$

where $x_t$ denotes the latent state and $u_t$ is the input token. The transition matrices $\bar{A} = \exp\big(\Delta A(u_t)\big)$ and $B(u_t)$ are dynamically computed, allowing spatially variant memory updates without explicit positional encoding. To operationalize this within 3D medical volumes, we adopt the VSS3D block from MedSegMamba[29], which routes directional scan sequences through stacked Mamba layers and modulates them via a lightweight MLP. Each block applies pre-normalization with LayerNorm, followed by 3D selective scanning (SS3D) that unfolds the volume along multiple anatomical axes. The inner residual connections preserve the original representation across the update path, while SiLU activations introduce smooth nonlinearity and DropPath regularization enhances robustness. This architecture captures both fine-grained local context and non-local continuity while remaining computationally efficient.

To benchmark the generalizability of our proposed multitask framework across different backbone types, we also implemented a vision Transformer[32] as a comparative baseline. The Transformer design introduces hierarchical attention with positional encoding to better capture cross-regional dependencies, and has been widely accepted in medical image synthesis literature[32,47]. This comparative setup allows us to assess the extent to which our multitask formulation improves performance across diverse architectures and highlights the plug-and-play compatibility of our framework with state-of-the-art vision models.

### 2.3. Implementation and Evaluation

Although the public dataset consists of 37 paired MRI/CT subjects, model training is performed using a 3D patch-based strategy rather than subject-level volumes, thereby alleviating the limitation imposed by the modest number of subjects. Specifically, we adopted a randomized patch sampling strategy, extracting 100 patches per subject from the training and validation sets at each epoch, respectively. Patch locations are randomly sampled under specific spatial constraints, such that the sampled patches vary across epochs. This yielded a total of 3,000 training samples and 300 validation samples, effectively expanding the dataset by two orders of magnitude. To enhance the model’s focus on bone structures, all patches used for voxel-wise HU regression were sampled around the skull center. In contrast, the segmentation task required broader foreground–background discrimination; thus, 80% of segmentation patches were sampled near the skull center, while the remaining 20% were drawn from peri-cranial and intracranial regions to increase anatomical diversity.

The multitask Res-Mamba U-Net was initially trained on a 1.5T public dataset and subsequently adapted to the 7T clinical data using transfer learning. Specifically, the full network (including the shared backbone and task-specific heads) was initialized with pretrained 1.5T weights and fine-tuned end-to-end on the 7T dataset using a reduced learning rate, without freezing any layers. The 7T dataset comprised 10 paired MRI/CT subjects and was used exclusively for fine-tuning and evaluation under a subject-level split. This strategy not only accelerates convergence and alleviates the need for extensive hyperparameter search, but also consistently improves performance in the target domain[48,49]—yielding higher accuracy for both skull segmentation (average Dice of sCT skull in the multitask network increased from 0.825 to 0.855, $p < 0.05$) and HU value prediction (average MAE of sCT skull in the multitask network decreased from 223.056 to 189.737, $p < 0.05$). Detailed results are provided in Supplementary Table 2.

For comparative evaluation, we preserved the single-task sCT model trained under the same conditions, with all factors controlled except for the network design itself. Specifically, both models were trained using the same paired MRI/CT subjects, and the same patch-based sampling and training strategy. All experiments were conducted under identical software and hardware environments, and for the 7T experiments, the single-task baseline was adapted using the same transfer learning procedure as the multitask model, allowing us to assess the added value of our multitask formulation. All training and validation procedures were conducted separately on the respective subsets of the two repositories. All benchmarking experiments were performed on the same NVIDIA A100 GPU under an identical software environment. Compared with the single-task baseline, the multitask framework introduced only lightweight task-specific decoder heads, increasing the total number of model parameters by approximately 15–20% without substantially altering the computational profile of the shared encoder. Under the same early-stopping criteria, the multitask model typically reached stable convergence within approximately 20–25 epochs, whereas the single-task model generally required around 40 epochs to achieve comparable convergence. As a result, the overall training burden remained comparable, while the multitask formulation also reduced reliance on additional post-processing by producing more anatomically consistent outputs directly.

To comprehensively evaluate the performance of the proposed method, we used a series of metrics as quantitative criteria, encompassing spatial correlation, morphological segmentation accuracy, and overall image quality assessment. Spatial correlation was evaluated using both Pearson and Spearman correlation coefficients, providing insights into linear and monotonic relationships, respectively. For assessing the accuracy of skull contour morphology segmentation, we utilized Dice and Jaccard coefficients, which are robust measures of spatial overlap. Overall image quality was gauged using multiple complementary metrics. The Structural Similarity Index (SSIM) was employed to evaluate the perceived quality of the generated images by comparing local patterns of pixel intensities. To quantify the fidelity of the reconstructed images, we utilized the Peak Signal-to-Noise Ratio (PSNR). Additionally, the Mean Absolute Error (MAE) was calculated to assess pixel-level prediction accuracy, providing a direct measure of the average magnitude of discrepancies between predicted and ground truth values. These evaluations were conducted on independent test sets derived from both datasets.

To ensure unbiased evaluation of image quality in sCT synthesis—particularly within skull regions—we applied thresholding and morphological masking to remove spurious signals in foreground and background areas from the single-task sCT predictions. The morphological mask was generated by applying a voxel-wise dilation to a group-level mean CT-derived bone template. In contrast, the proposed multitask network required no additional post-processing, as anatomical structure learning was explicitly integrated into the synthesis process.

## 3. RESULTS

Fig. 2 illustrates the experimental results obtained on 1.5T public brain dataset. Residual maps were generated by subtracting the original CT from the sCT produced under the respective processing framework and subsequently rescaling the values. All visualized images are presented along three anatomical planes. Compared to the single-task Mamba U-Net, our multitask model yielded visibly sharper structural delineation and more realistic HU distributions, particularly along complex bone contours, highlighting the synergy between structural awareness and voxel-level accuracy. Beyond enhanced boundary precision, the multitask network also substantially reduced foreground and background noise in the generated sCT images, as evidenced by the binary mask comparison. Because conventional methods struggle to preserve soft-tissue contrast effectively, even with post-processing using threshold and morphological masks, the peak signal-to-noise ratio of the sCT skull image is still inferior to that achieved by the multitask network, as shown in Table 1 (average PSNR 27.614 in single-task network vs. 28.898 in multitask, $p < 0.05$). This underscores the method's robustness to artifacts and its contribution to overall image-quality stability.

Table 1 summarizes the quantitative performance of models in different tasks on the 1.5T test dataset. Statistical significances were assessed using paired t-tests. Across all four test subjects, the multitask MambaU consistently outperformed the single-task counterpart, in terms of image correlation, segmentation accuracy, and overall quality (all $p < 0.05$). Especially for morphological relationship, the multitask network improved the results significantly (average Dice score increased from 0.885 to 0.927, average Jaccard index from 0.795 to 0.864, all $p < 0.01$), demonstrating the potential contribution to ROI delineation using a dedicated segmentation network task. Enhanced regional differentiation is evident at the voxel level. With multitask learning, bone-specific MAE declined by 4.3% compared with the template-averaged single-task model ($p < 0.05$), while whole-brain MAE decreased by 44.4%, from 121.8 to 67.7 HU ($p < 0.001$). Although the relative reduction within bone appears modest, the marked global improvement indicates that optimizing high-HU bone regions—often underrepresented in global loss functions—provides a decisive contribution to overall synthesis quality. Conversely, whole-brain averaging dilutes residual bone errors with low-HU soft tissue, thereby overestimating model performance. These findings emphasize the importance of bone-aware training and region-specific evaluation for an unbiased assessment of sCT quality.

To further assess generalizability, we fine-tuned the pre-trained model obtained from a 1.5T public repository on a 7T clinical cohort. The consistent results (Fig. 3) provide evidence that the model remains robust even under more extreme imaging conditions such as 7T MRI. Fig. 3a-d presents a representative case from the clinical dataset. In addition to the disparities among field strength and sequence parameters compared to the original training dataset, the new T1w MP2RAGE images exhibit more reconstruction artifacts and lead to lower prediction PSNR, particularly in the cavity regions adjacent to the skull tissue. These artifacts impair reliable bone structure extraction and exacerbate morphological ambiguity. Consequently, the single-task Mamba U-Net exhibited substantial degradation in performance under direct MRI-to-CT synthesis, as reflected in Fig. 3f, 3h and the residual visualization in Fig. 3j. Without anatomical guidance, MSE-driven optimization led to severe morphological distortion in the predicted sCT, including spurious soft tissue intensities and background noise—rendering the output clinically unreliable even after conventional post-processing steps such as thresholding or group-averaged masking. In contrast, the multitask network results (Fig. 3e, 3g, and residual in 3i) achieved high-quality synthesis with preserved skull morphology and minimal artifacts, without requiring any post-processing, underscoring the benefit of joint anatomical supervision under challenging acquisition conditions.

The importance of the segmentation task is further corroborated by the quantitative results presented in Fig. 3k. Across all evaluation metrics, the multitask network achieved significant improvements over the single-task approach (all $p < 0.001$, except for PSNR, $p = 0.005$), surpassing even the improvements observed in the 1.5T dataset, e.g. average Dice score increases by 12.6% (from 0.759 to 0.855) in 7T vs. 4.7% (from 0.885 to 0.927) in 1.5T respectively. Detailed results are provided in Supplementary Table 1. This disparity reflects not only input-quality differences but also the inherent challenges of clinical acquisition and standardization. Unlike research datasets, clinical scans are subject to greater variability in protocol, hardware, and patient motion, often lacking consistent registration pipelines or curated anatomical templates. As a result, post-processing strategies such as group-averaged CT masking become unreliable, weakening boundary localization in single-task models. These findings highlight the limitations of conventional pipelines in real-world clinical deployment and underscore the robustness advantage conferred by task-structured anatomical supervision.

To further assess the contribution of segmentation accuracy, we incorporated ground-truth CT-derived masks as reference inputs during image quality evaluation, as indicated by the darker bars in Fig. 3k. Using identical HU prediction model parameters, the incorporation of more precise anatomical masks led to consistently higher PSNR and lower MAE, illustrating the benefit of accurate spatial guidance. While such oracle masks are not available during inference and serve only as an upper-bound estimate, this finding reinforces the critical role of the segmentation pathway within our multitask framework.

As an additional investigation into the plug-and-play flexibility and backbone-agnostic robustness of our cascaded multitask framework, we performed a comparative analysis between TransU and MambaU on the 7T dataset, as illustrated in Fig. 4. As shown in the boxplots (Fig. 4a–g), MambaU consistently outperformed TransU across nearly all evaluation metrics in both single-task and multitask configurations. While these differences support the superior modeling capacity of the Mamba-based spatial state-space architecture, the relative performance gains shown in Fig. 4h underscore a more notable enhancement effect on the weaker backbone. The multitask formulation significantly boosted TransU's performance across all evaluated metrics, nearly closing the gap with MambaU in several key domains. For instance, segmentation performance under the multitask setting showed near-identical Dice scores (TransU: 0.824 vs. MambaU: 0.855) and similarly convergent SSIM values (TransU: 0.908 vs. MambaU: 0.927). In terms of relative gains over the single-task baselines, Dice improved by 50.4% for TransU and 12.9% for MambaU, while SSIM increased by 13.8% and 4.0%, respectively (as shown in Fig. 4h). Detailed results are provided in Supplementary Table 3 and 4. This convergence highlights the framework's ability to bridge architectural disparities, enabling models with moderate baseline performance to benefit from structurally guided supervision and joint optimization. Nevertheless, noticeable gaps remained in image detail metrics such as PSNR and MAE, suggesting that the performance alignment is primarily driven by reduced segmentation differences. These more anatomically consistent attention masks—produced even by the benchmark model—appear sufficient to guide collaborative HU prediction and ultimately enhance the fidelity of the synthesized CT outputs.

## 4. DISCUSSION

Our proposed multitask framework introduces a structural advance in MRI-to-CT modality conversion by incorporating skull segmentation as an anatomically grounded precursor to HU regression. This cascaded design enables segmentation-derived attention masks to dynamically guide region-specific loss computation for HU prediction, embedding anatomical priors to enhance spatial specificity and clinical relevance. Compared to conventional direct regression models that rely solely on pixel-level supervision, our formulation offers both computational advantages through localized loss weighting and improved predictive accuracy in bone-relevant regions. Notably, the superiority of this framework arises from methodological innovation rather than backbone complexity. Experimental comparisons demonstrate that multitask learning consistently improves performance across all tested architectures and substantially narrows the performance gap between strong (MambaU) and weaker (TransU) backbones. By leveraging structure-guided supervision, benchmark models can approach the performance of more advanced architectures in critical subregions. Moreover, this study represents the first application of Mamba-based state-space modeling to CT synthesis, validating its effectiveness in capturing long-range spatial dependencies in high-resolution 3D medical imaging.

This strategy demonstrates robust generalizability under clinically relevant variability, where single-task models, even after fine-tuning, often degrade when exposed to heterogeneous inputs such as 7T clinical scans with altered imaging parameters. The main limitations arise from non-standard acquisition protocols, inconsistent spatial alignment, and weak foreground–background contrast. In contrast, our multitask approach maintains robust performance without additional post-processing by leveraging task coupling and anatomical priors to adapt across scanners and field strengths. Although 7T MRI intrinsically provides higher SNR and resolution, its clinical datasets are far more heterogeneous[50] than the standardized 1.5T data, with diverse protocols and substantial inter-subject variation that complicate sCT generation. These difficulties are further compounded by field-specific factors such as B0/B1 inhomogeneity[51], susceptibility artifacts, and markedly shorter T2* values, all of which weaken bone-to-air contrast. Nevertheless, by embedding structural priors and coupled supervision, our framework achieves stable outputs even under the challenging conditions of real-world 7T clinical imaging.

Additionally, the use of 3D volume training and evaluation throughout this study eliminates slice-level bias common in prior 2D-based sCT approaches. By embedding training within a 3D patch extraction strategy, the framework avoids the need for manual image resizing, reduces the computational burden of volumetric inputs, and enables the model to focus on localized anatomical structure. In contrast to previous studies, our approach achieves promising results on limited datasets without requiring additional data augmentation. Nonetheless, patch size remains a critical factor influencing model performance: overly small patches may fail to capture global context, while large stride values in patch sampling can introduce reconstruction artifacts. In practice, potential "border effects" during patch-based reconstruction can be mitigated by reducing the stride size, thereby preserving structural continuity without inflating the sample count.

While the proposed framework demonstrates robust performance under heterogeneous MRI acquisition conditions, the present study is best interpreted as a methodological validation. Its clinical relevance lies in supporting prerequisite advances for synthetic CT workflows, particularly by improving robustness across heterogeneous MRI settings and enhancing skull-related fidelity in bone-sensitive applications. Potential scenarios include MRI-only or CT-sparse pipelines, non-standard or ultra-high-field MRI protocols, and downstream applications such as radiotherapy planning, PET attenuation correction, and transcranial focused ultrasound. Nevertheless, the current evaluation is limited to the cranial region and to non-oncologic clinical brain imaging data, without radiotherapy-specific anatomy, treatment planning information, electron density assessment, or dosimetric validation. The findings therefore establish technical feasibility only within this defined scope. Clinical translation will require validation in oncologic cohorts, incorporation of treatment-planning datasets, quantitative electron-density analysis, clinically established dosimetric endpoints such as dose difference, DVH, and gamma analysis, and extension to additional anatomical sites through broader multi-center evaluation.

## 5. CONCLUSION

In this study, we proposed a cascaded multitask learning framework for multimodal MRI-to-CT synthesis. Validated across datasets of varying field strengths and acquisition settings, our results demonstrate that the proposed approach substantially enhances both predictive performance and generalizability compared to conventional single-task synthetic CT generation.

A key contribution of this work lies in the comparative analysis of sCT generation across heterogeneous data sources, with a focused evaluation of the cranial region. By structurally decoupling the task into skull segmentation and region-constrained HU regression, and isolating osseous structures from soft tissue, our framework enables more targeted and clinically relevant assessment of image quality. This methodological design provides clearer insights into model behavior under realistic imaging conditions and highlights the value of structured, anatomy-aware learning for addressing key methodological challenges in cranial sCT generation.

## ETHICS STATEMENT

This study utilized both publicly available and institutionally acquired human brain datasets. The 1.5T data were obtained from an open-access repository and were used in accordance with the ethical regulations and data use policies of the original data provider. The 7T dataset comprises clinically acquired imaging data collected under institutional protocols and handled in compliance with applicable confidentiality requirements.

## DATA AVAILABILITY

The 1.5T dataset used in this study is publicly available from the original open-access repository. The 7T dataset is not publicly available because it contains institutionally acquired clinical imaging data subject to confidentiality restrictions. Access to this dataset may be considered upon reasonable request to the corresponding authors and with permission from the relevant institution.

## AUTHOR CONTRIBUTIONS

Z.X. conceived and designed the study, implemented the 3D-ResMamba multitask framework, and conducted the experiments and analyses. Y.Z. contributed to model training, figure preparation, and data processing. E.G. and E.H.M. were responsible for data acquisition and preprocessing. J.H. and J.G. supervised the project, provided guidance on study design and interpretation, and critically revised the manuscript. All authors discussed the results and approved the final version of the manuscript.

## CONFLICT OF INTEREST

The authors declare no conflicts of interest.

## FUNDING

This research was supported by the National Institutes of Health through grants from the National Institute of Neurological Disorders and Stroke (NINDS; 1R01NS108452 and 1R01NS120879), the National Institute on Aging (NIA; 5R01AG064093 and 1RF1AG082257), and the National Institute of Biomedical Imaging and Bioengineering (NIBIB; P41 EB031771).

## ACKNOWLEDGEMENTS

This work also made use of the CERMEP-IDB-MRXFDG Database (© Copyright CERMEP – Imagerie du vivant and Hospices Civils de Lyon. All rights reserved.), provided jointly by CERMEP and Hospices Civils de Lyon (HCL) under a free academic end-user license agreement.

## FIGURE AND TABLE LEGENDS

**Table 1.** Quantitative evaluation results of MambaU synthetic CT based on 1.5T public dataset.

| Index | Correlation | | Segmentation | | Overall quality | | | |
|---|---|---|---|---|---|---|---|---|
| | Pearson | Spearman | Dice | Jaccard | SSIM | PSNR | MAE (Bone) | MAE (Brain) |
| Single-Task | | | | | | | | |
| Sub1 | 0.961 | 0.923 | 0.920 | 0.851 | 0.956 | 27.140 | 165.947 | 128.858 |
| Sub2 | 0.916 | 0.866 | 0.863 | 0.760 | 0.943 | 26.852 | 185.353 | 122.834 |
| Sub3 | 0.952 | 0.893 | 0.889 | 0.801 | 0.957 | 29.262 | 132.279 | 114.619 |
| Sub4 | 0.937 | 0.873 | 0.869 | 0.768 | 0.947 | 27.203 | 174.607 | 121.068 |
| Avg | 0.942 | 0.889 | 0.885 | 0.795 | 0.951 | 27.614 | 164.547 | 121.845 |
| Multi-Task | | | | | | | | |
| Sub1 | 0.962 | 0.929 | 0.942 | 0.891 | 0.960 | 27.454 | 158.084 | 80.174 |
| Sub2 | 0.923 | 0.897 | 0.908 | 0.832 | 0.952 | 27.640 | 178.048 | 63.828 |
| Sub3 | 0.962 | 0.931 | 0.941 | 0.888 | 0.968 | 30.882 | 124.882 | 68.751 |
| Sub4 | 0.943 | 0.906 | 0.916 | 0.844 | 0.955 | 28.616 | 168.748 | 57.972 |
| Avg | 0.948 | 0.916 | 0.927 | 0.864 | 0.959 | 28.898 | 157.441 | 67.681 |
| p-value | <0.05 | <0.05 | <0.01 | <0.01 | <0.05 | <0.05 | <0.05 | <0.001 |

* Pearson = Pearson's correlation, Spearman = Spearman's rank correlation, Dice = Sørensen–Dice coefficient, Jaccard = Jaccard similarity coefficient, SSIM = Structural similarity index, PSNR = Peak Signal-to-Noise-Ratio, MAE = Mean Absolute Error, HU = Hounsfield Unit.

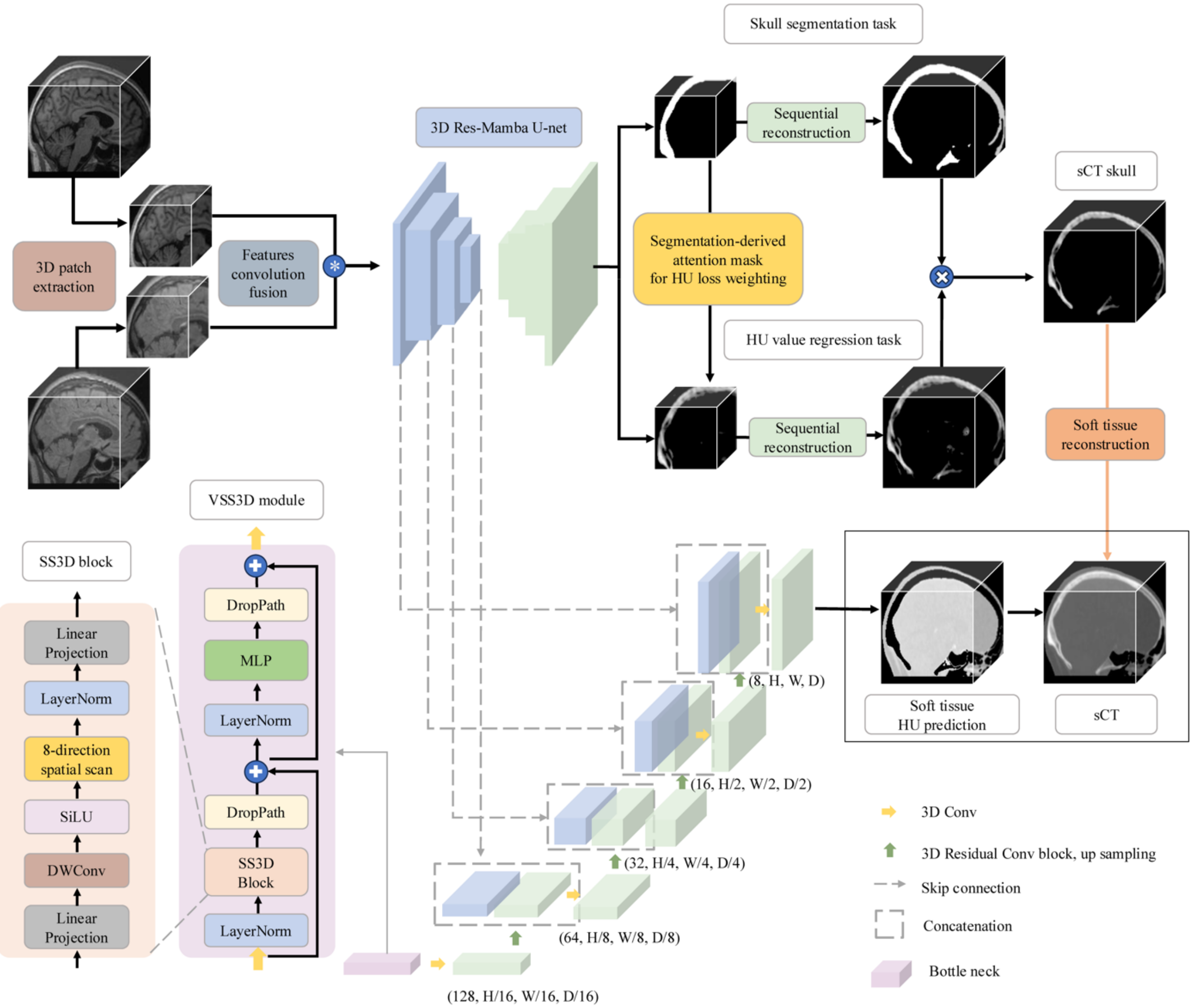

**Fig. 1.** Proposed 3D Res-Mamba U-Net–based multitask processing framework for sCT generation. The pipeline integrates dual MRI inputs (T1w and T2-FLAIR) and performs 3D patch extraction before entering a shared encoder–decoder backbone. A residual Mamba-based state space module is embedded at the bottleneck to enhance long-range dependency modeling. The framework cascades two coupled tasks: a skull segmentation branch, providing anatomical priors, and a region-constrained Hounsfield Unit (HU) regression branch, supervised by dynamic attention masks derived from the segmentation output. An auxiliary soft-tissue regression branch complements bone-focused HU prediction, and outputs from all branches are fused to reconstruct the final sCT volume.

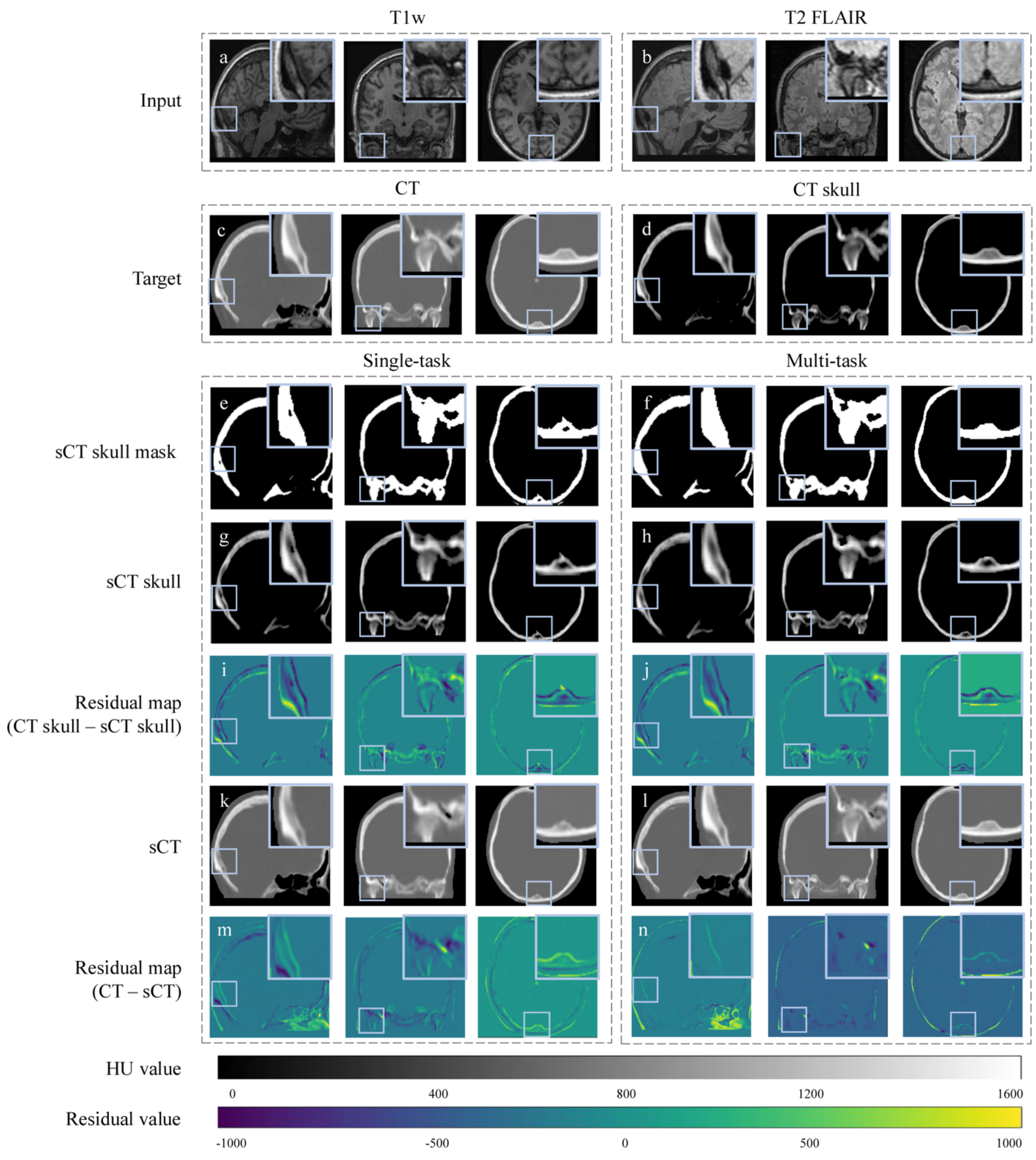

**Fig. 2.** Visualization of 1.5T dataset and synthetic CT results across three anatomical planes. (a) Original T1w MRI, (b) T2 FLAIR MRI, and (c) real CT images and (d) real CT skull images, shown in sagittal, coronal, and transverse planes. (e) Skull segmentation mask of synthetic CT images using direct ResMambaUNet prediction, with post-processing of thresholding and morphological mask. (f) Skull segmentation mask of sCT generated from multitask networks. (g)(h) Synthetic CT images based on T1-weighted and T2 FLAIR MRI inputs, using single-task and multitask networks, respectively. (i)(j) The corresponding residual maps (real CT skull - sCT skull) for the synthetic CT results in HU. (k,l) Synthetic CT images (sCT) from single-task and multitask networks, respectively. (m,n) Corresponding residual maps (real CT – sCT) for the single-task and multitask networks.

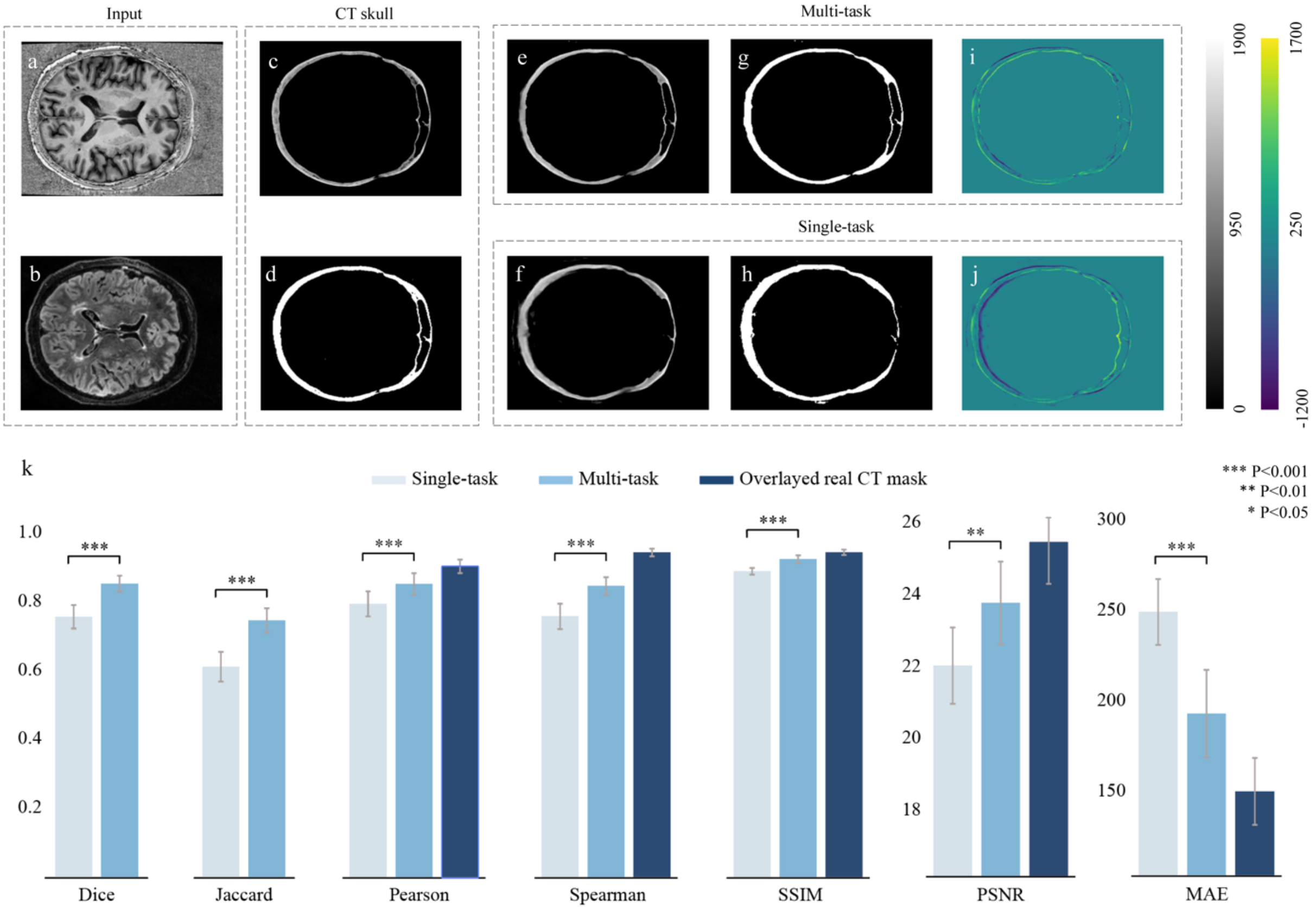

**Fig. 3.** Visualization of 7T clinical dataset sample, corresponding synthetic CT skull images, and evaluation results. (a) original images of T1-weighted MP2RAGE, (b) T2 FLAIR, (c) CT skull with soft tissue removed, and (d) CT skull mask generated from binarization of (c). (e) Skull segmentation mask of sCT generated from multitask networks. (f) Skull segmentation mask of synthetic CT images using direct MambaU prediction, with post-processing of thresholding and morphological mask. (g)(h) Synthetic CT images based on T1-weighted MP2RAGE and T2 FLAIR MRI inputs, using multitask and single-task networks, respectively. (i)(j) The corresponding residual maps (real CT skull – sCT skull) for the synthetic CT results in HU. (k) Visualization of the evaluation metrics corresponding to sCT from single-task network with postprocessing, multitask network, and direct HU prediction overlaid with real CT mask as a reference.

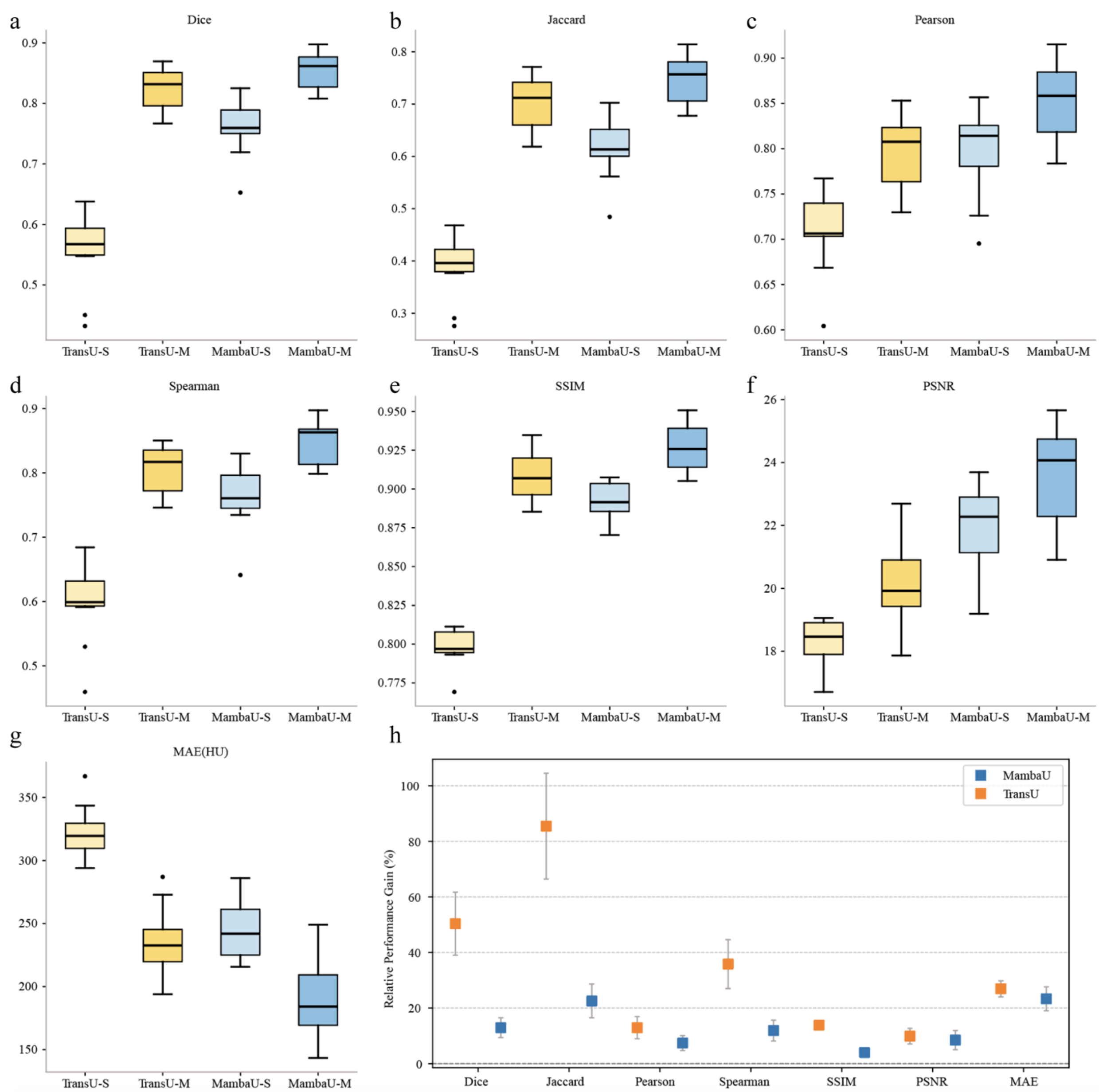

**Fig. 4.** Quantitative performance comparison of multitask and single-task networks across two model architectures. (a–g) Boxplots of evaluation metrics on the 7T clinical dataset comparing single-task (–S) and multitask (–M) configurations of TransU (yellow) and MambaU (blue). The assessed metrics include Dice coefficient (a), Jaccard index (b), Pearson correlation (c), Spearman correlation (d), structural similarity index (SSIM) (e), peak signal-to-noise ratio (PSNR) (f) and mean absolute error (MAE) in Hounsfield Units (HU) (g). (h) Summary plot of the relative performance gain (%) from multitask training over single-task baselines for each metric. Dots indicate mean gain with standard deviation across test subjects. Orange and blue denote TransU and MambaU models, respectively.

**SUPPLEMENTARY MATERIALS**

## Table of Contents

**SUPPLEMENTARY TABLES**

## Supplementary Table S1

Performance comparison of single-task and multitask Mamba-UNet models on the 7T test set.

| Index | Mamba-U (single-task) | | | Mamba-U (multi-task) | | | *p*-value | Relative increase (%) † |
|---|---|---|---|---|---|---|---|---|
| | Mean | CI lower | CI upper | Mean | CI lower | CI upper | | |
| Correlation | | | | | | | | |
| Pearson | 0.797 | 0.755 | 0.838 | 0.854 | 0.818 | 0.890 | <0.001 | 7.245 |
| Spearman | 0.760 | 0.718 | 0.802 | 0.849 | 0.819 | 0.878 | <0.001 | 11.607 |
| Segmentation | | | | | | | | |
| Dice | 0.759 | 0.720 | 0.798 | 0.855 | 0.829 | 0.882 | <0.001 | 12.670 |
| Jaccard | 0.614 | 0.565 | 0.663 | 0.748 | 0.708 | 0.788 | <0.001 | 21.900 |
| Overall quality | | | | | | | | |
| SSIM | 0.891 | 0.881 | 0.902 | 0.927 | 0.914 | 0.940 | <0.001 | 3.961 |
| PSNR | 21.874 | 20.661 | 23.086 | 23.619 | 22.308 | 24.930 | <0.001 | 7.981 |
| MAE (HU) | 246.012 | 225.253 | 266.770 | 189.737 | 162.074 | 217.401 | <0.001 | 22.875 |

* “Single-task” denotes the baseline Mamba-UNet trained for skull-only synthetic CT prediction on the 7T dataset. “Multi-task” denotes the multitask Mamba-UNet jointly trained on cascaded skull segmentation task and skull HU value regression task on the 7T dataset.

† Relative increase (%) was calculated as the percentage difference in mean performance between the multitask and single-task Mamba-U models.

**Supplementary Table S2**

Evaluation of multitask Mamba-UNet before and after transfer learning on the 7T test set.

| Index | Mamba-U (pre-TL) | | | Mamba-U (post-TL) | | | *p*-value | Relative increase (%) † |
|---|---|---|---|---|---|---|---|---|
| | Mean | CI lower | CI upper | Mean | CI lower | CI upper | | |
| Correlation | | | | | | | | |
| Pearson | 0.818 | 0.778 | 0.858 | 0.854 | 0.818 | 0.890 | 0.002 | 4.389 |
| Spearman | 0.817 | 0.779 | 0.855 | 0.849 | 0.819 | 0.878 | <0.001 | 3.822 |
| Segmentation | | | | | | | | |
| Dice | 0.825 | 0.791 | 0.860 | 0.855 | 0.829 | 0.882 | <0.001 | 3.609 |
| Jaccard | 0.705 | 0.655 | 0.754 | 0.748 | 0.708 | 0.788 | <0.001 | 6.172 |
| Overall quality | | | | | | | | |
| SSIM | 0.915 | 0.901 | 0.928 | 0.927 | 0.914 | 0.940 | <0.001 | 1.343 |
| PSNR | 22.573 | 21.620 | 23.526 | 23.619 | 22.308 | 24.930 | 0.023 | 4.637 |
| MAE (HU) | 223.056 | 196.847 | 249.265 | 189.737 | 162.074 | 217.401 | <0.001 | 14.937 |

* “Pre-TL” denotes the baseline multitask Mamba-UNet directly trained from scratch on the 7T dataset. “Post-TL” denotes the transfer-learned Mamba-UNet that was first pretrained on the 1.5T dataset for synthetic CT skull prediction, and subsequently fine-tuned on the 7T dataset.

† Relative increase (%) was calculated as the absolute percentage difference between the mean performances of the post-TL and pre-TL models.

**Supplementary Table S3**

Performance comparison of single-task and multitask Transformer-UNet models on the 7T test set.

| Index | Transformer-U (single-task) | | | Transformer-U (multi-task) | | | *p*-value | Relative increase (%) † |
|---|---|---|---|---|---|---|---|---|
| | Mean | CI lower | CI upper | Mean | CI lower | CI upper | | |
| Correlation | | | | | | | | |
| Pearson | 0.708 | 0.669 | 0.746 | 0.798 | 0.763 | 0.832 | <0.001 | 12.680 |
| Spearman | 0.598 | 0.546 | 0.651 | 0.805 | 0.773 | 0.838 | <0.001 | 34.593 |
| Segmentation | | | | | | | | |
| Dice | 0.555 | 0.501 | 0.609 | 0.824 | 0.794 | 0.854 | <0.001 | 48.500 |
| Jaccard | 0.387 | 0.337 | 0.437 | 0.702 | 0.658 | 0.747 | <0.001 | 81.475 |
| Overall quality | | | | | | | | |
| SSIM | 0.798 | 0.788 | 0.808 | 0.908 | 0.896 | 0.921 | <0.001 | 13.815 |
| PSNR | 18.250 | 306.260 | 339.888 | 20.065 | 18.937 | 21.193 | <0.001 | 9.946 |
| MAE (HU) | 323.074 | 196.847 | 249.265 | 236.608 | 214.344 | 258.873 | <0.001 | 26.763 |

* "Single-task" denotes the baseline Transformer-UNet trained for skull-only synthetic CT prediction on the 7T dataset. "Multi-task" denotes the multitask Transformer-UNet jointly trained on cascaded skull segmentation task and skull HU value regression task on the 7T dataset.

† Relative increase (%) was calculated as the percentage difference in mean performance between the multitask and single-task Transformer-U models.

**Supplementary Table S4**

Relative performance gains of Transformer-U and Mamba-U on the 7T test set.

| Index | Transformer-U | | | Mamba-U | | |
|---|---|---|---|---|---|---|
| | Mean (%) | CI lower (%) | CI upper (%) | Mean (%) | CI lower (%) | CI upper (%) |
| Correlation | | | | | | |
| Pearson | 12.936 | 8.932 | 16.940 | 7.399 | 4.716 | 10.082 |
| Spearman | 35.847 | 27.060 | 44.634 | 11.905 | 8.148 | 15.662 |
| Segmentation | | | | | | |
| Dice | 50.539 | 39.007 | 61.711 | 12.933 | 9.388 | 16.478 |
| Jaccard | 85.457 | 66.411 | 104.503 | 22.567 | 16.476 | 28.658 |
| Overall quality | | | | | | |
| SSIM | 13.826 | 12.662 | 14.990 | 3.958 | 3.582 | 4.334 |
| PSNR | 9.896 | 7.084 | 12.708 | 8.456 | 5.017 | 11.895 |
| MAE (HU) | 26.923 | 24.029 | 29.817 | 23.327 | 19.034 | 27.620 |

* Relative gain was calculated for each subject as the relative percentage change between single-task and multitask performances. Group-level estimates were summarized by the mean relative gain, and the 95% confidence interval was obtained as mean ± 1.96 × SE, where SE denotes the standard error of the mean.